\documentclass[journal=ancac3,manuscript=article]{achemso}

\usepackage[normalem]{ulem}
\usepackage[usenames, dvipsnames]{color}
\usepackage{dcolumn}
\usepackage{amsmath}
\usepackage{adjustbox}
\usepackage{abstract}
\usepackage{multirow}

\author{E. G. Marin}
\affiliation[Universita' di Pisa]
{Dipartimento di Ingegneria dell'Informazione, Universit\`{a} di Pisa, 56122, Pisa, Italy}
\alsoaffiliation[Universidad de Granada]
{Dpto. Electr\'onica, Fac. Ciencias, Universidad de Granada, 18071, Granada, Spain}
\author{D. Marian}
\affiliation[Universita' di Pisa]
{Dipartimento di Ingegneria dell'Informazione, Universit\`{a} di Pisa, 56122, Pisa, Italy}
\author{M. Perucchini}
\affiliation[Universita' di Pisa]
{Dipartimento di Ingegneria dell'Informazione, Universit\`{a} di Pisa, 56122, Pisa, Italy}

\author{G. Fiori}
\affiliation[Universita' di Pisa]
{Dipartimento di Ingegneria dell'Informazione, Universit\`{a} di Pisa, 56122, Pisa, Italy}
\author{G. Iannaccone}
\affiliation[Universita' di Pisa]
{Dipartimento di Ingegneria dell'Informazione, Universit\`{a} di Pisa, 56122, Pisa, Italy}
\email{giuseppe.iannaccone@unipi.it}

\title{Lateral Heterostructure Field-Effect Transistors Based on 2D-Material Stacks With Varying Thickness and Energy Filtering Source}

\begin{document}
	
	\maketitle


\begin{abstract}	
  The bandgap dependence on the number of atomic layers of some families of 2D-materials, can be exploited to engineer and use lateral heterostructures (LHs) as high-performance Field-Effect Transistors (FET). This option can provide very good lattice matching as well as high heterointerface quality. More importantly, this bandgap modulation with layer stacking can give rise to steep transitions in the density of states (DOS) of the 2D material, that can eventually be used to achieve sub-$60$ mV/decade subthreshold swing in LH-FETs thanks to an energy-filtering source. We have observed this effect in the case of a PdS$_2$  LH-FET due to the particular density of states of its bilayer configuration. Our results are based on \latin{ab initio} and multiscale materials and device modeling, and incite the exploration of the 2D-material design space in order to find more abrupt DOS transitions and better suitable candidates. 
\end{abstract}

\newpage


Semiconductor heterostructures of the III-V and II-VI materials systems have played a fundamental role in the progress of electronics and optoelectronics. Firstly proposed by Kroemer in the 1950s \cite{Kroemer01}, they have been involved in the invention of  
quantum-well lasers \cite{Ziel1975} and high-electron-mobility transistors \cite{Mimura80}.
The large number of available two-dimensional (2D) materials and the possibility to combine them even in the presence of significant lattice mismatch has led to a new wave of interest in materials engineering based on heterostructures of 2D materials. In particular, 2D materials enable the realization of vertical heterostructures, also called 
``van der Waals" heterostructures, consisting in the vertical stacking of layers of different 2D materials loosely coupled by van der Waals interactions \cite{Geim13,Novoselov16}, and of lateral heterostructures (LHs), in which a single 2D layer consists of juxtaposed regions of different lattice-matched 2D materials \cite{Ci10,Levendorf12,Liu13,Huang2014lateral}.

LHs have been shown to be particularly well suited as channel materials in high performance Field-Effect Transistors (FETs) for digital electronics \cite{Iannaccone18}.
However, the quality of the heterojunction is one of the major obstacles towards the experimental demonstration of high performance LH-FETs. The possibility of fabricating LHs by modulating the stacking order of a single 2D material provides the opportunity of perfect lattice matching and growth compatibility, and therefore a chance to obtain high materials quality \cite{Ghorbani16}. 

Recently, a particular group of transition metal dichalcogenides (TMDs) involving noble transition metals (Pt, Pd, and Ni), combined with S, Se, and Te, have been predicted  \cite{Wang15b} and demonstrated to have strong gap dependence on the number of stacked layers \cite{Ciarrocchi18}. The so-called ``noble TMDs" are, thus, promising contenders to build 2D LHs by modulation of the number of layers of adjacent regions of the same material. Indeed, these structures based on noble TMDs --that strictly speaking could be considered homostructures instead of heterostructures-- would be easier to realize than those made of different 2D materials. \latin{Ab initio} calculations predict that the bandgap is reduced of more than 1~eV when these noble TMDs vary from monolayer (1L) to bilayer (2L), leading in some cases to a change of electronic phase from semiconductor to metal \cite{Miro14b, Wang15b}. 
Monolayer and few-layer PtS$_2$ and PtSe$_2$ have already been synthesized  \cite{Zhao16,Wang15,Yan17}, and devices such as Schottky barrier diodes on silicon have been fabricated \cite{Yim16}. More recently, FETs made of few-layer PdSe$_2$ and PtSe$_2$ have  been experimentally realized showing ambipolar transfer characteristics \cite{Chow17,Zhao17,Ciarrocchi18} and a large dependence of PtSe$_2$ conductance on the number of layers has been observed \cite{Ciarrocchi18}.

There is an additional advantage of some noble TMDs in their bilayer form: their density of states (DOS) exhibits a steep non-monotonic variation around the Fermi energy that can be used as an energy filtering mechanism in order to obtain a subthreshold swing (SS) of the FET smaller than the Boltzmann limit of $60$~mV/decade at room temperature. Indeed, in thermionic FETs the inverse of the maximum slope of the current-voltage characteristics is limited to SS$ = k_BT/q \ln(10)$ per decade, where $k_B$ is Boltzmann constant, $q$ is the elementary charge, and $T$ is the absolute temperature \cite{Qiu18,Liu18,logoteta2019steep}. This provides a room temperature SS value of $=60$~mV/decade. A lower value can only be obtained if 1) injection is energy-constrained, such as in a Tunnel FET \cite{Appenzeller2004,Seabaugh10} or using impact ionization \cite{Gopalakrishnan2002}, or 2) if an effective negative capacitance is realized in the gate oxide stack, thus amplifying the surface potential at the channel \cite{Salahuddin2008,Salvatore2008}. In particular, the energy-constrained injection from a steep DOS source has been demonstrated very recently using a graphene source in combination with a CNT channel in \cite{Qiu18} or a MoS$_2$ channel in \cite{Liu18,logoteta2019steep}. We show here that this effect can be achieved in a system based on a single 2D material, specifically PdS$_2$, where a bilayer source (injecting into a monolayer channel) can provide sufficient energy filtering to yield an SS below the Boltzmann limit.

In this work we investigate the potential of noble TMDs as channel materials for LH-FETs, showing the possibility to engineer an energy-filtering source in order to obtain sub-$60$~mV/dec SS at room temperature and to design devices with competitive figures of merit when compared to the predictions of the International Roadmap for Devices and Systems (IRDS) \cite{IRDS}. Finally, we show that noble TMDs enable the realization of a 2D resonant tunneling diode (RTD) based on LHs obtained with modulation of the number of atomic layers.

\section{Results and discussion}

\subsection{Lateral Heterostructure FETs}

We have selected those noble TMDs exhibiting a sharp non-monotonic DOS around the Fermi level and therefore potentially able to achieve SS $<60$~mV/decade at room temperature. Following bandstructure calculations available in the literature \cite{Miro14b,Ghorbani16}, we have opted for PdS$_2$ and NiS$_2$. Both materials are semimetal in bilayer form and semiconductors in monolayer form, so it is possible to build a FET with a lateral heterostructure formed by a bilayer source, a monolayer central channel region, and a bilayer drain. This results in two semimetal-semiconductor Schottky barriers at the source and drain ends. For the sake of clarity, a semimetal is a material with zero gap but a relatively low density of states around the Fermi energy.
We have also considered PtS$_2$ which presents a strong modulation of the bandgap from monolayer to bilayer, still maintaining a semiconducting gap.

\begin{figure}[t]
	\centering
	\includegraphics[width=1.0\columnwidth]{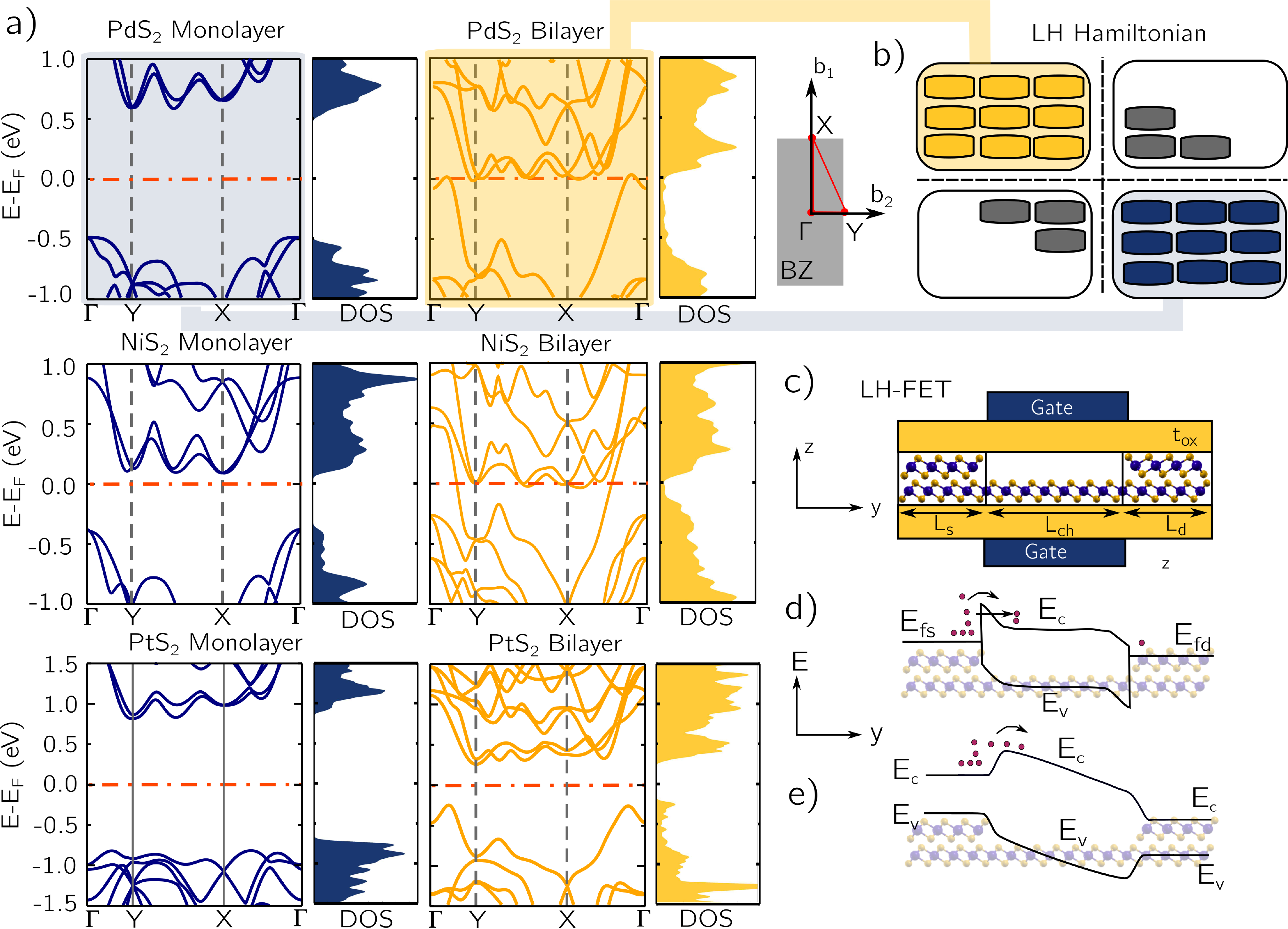}
	\caption{(a): Electronic band-structure on a highly symmetric path along the Brillouin zone depicted aside in gray with the path marked in red, and density of states integrated in the whole Brillouin zone, as computed with Density Functional Theory, for monolayer (1L) and bilayer (2L) PdS$_2$, NiS$_2$, and PtS$_2$. (b): Schematic  representation  of  the  construction  of the lateral heterostructure Hamiltonian from the Hamiltonian of the bilayer and monolayer materials. (c): Schematic of the Lateral Heterostructure FET, with bilayer source, monolayer channel, and bilayer-drain. (d) Illustration of the energy band edge profile of the  PdS$_2$ and NiS$_2$ LH-FETs , and (e) band edge profile of PtS$_2$ LH-FET.}
	\label{fig:Fig1}
\end{figure}

We have adopted a multi-scale simulation approach combining different levels of physical abstraction, ranging from \latin{ab initio} calculations of materials properties to full device simulations based on coherent quantum transport \cite{Marin18}. 
We have calculated the electronic band-structure of monolayer and bilayer PdS$_2$, PtS$_2$, and NiS$_2$ using density functional theory (DFT) as implemented in the Quantum Espresso suite~\cite{QE} (see Methods). The strong dependence of the electronic structure on the number of layers is highlighted in Fig. \ref{fig:Fig1}(a): when the crystal structure is varied from monolayer to bilayer, PdS$_2$ and NiS$_2$ undergo a phase change from semiconductor to semimetal, whereas PtS$_2$ has its energy gap reduced from 1.59~eV to 0.48~eV. The drastic variation of the DOS of bilayer PdS$_2$ and NiS$_2$ around the Fermi level can be exploited to inject carriers from the source 
with sub-maxwellian energy tails, leading to SS $<60$~mV/decade at room temperature.

\fboxrule=2pt
	
The plane-wave DFT basis set has been translated into a Maximally-Localized-Wannier-Functions (MLWFs) basis set by means of the Wannier90 code \cite{Wannier}, that provides us with tight-binding (TB) Hamiltonians for every material and stacking (see Methods and Fig. S1 in the Supplementary Information). The TB Hamiltonians are then employed to build a total Hamiltonian of the lateral heterostructure devices, following the procedure developed and validated in \cite{Marian17} (Fig. \ref{fig:Fig1}(b) and Methods). In order to accurately model the Schottky barrier formed at the bilayer/monolayer interfaces we have performed an energy analysis from first-principles taking into account the band offsets and the formation of dipoles \cite{Katagiri16} (see Supplementary Information).

The considered LH-FETs are also illustrated in Fig. \ref{fig:Fig1}(c). The length of the bilayer source and drain regions are $L_{\rm s/d}=11$~nm, $10.4$~nm and $16.7$~nm for PdS$_2$, NiS$_2$, and PtS$_2$ respectively. They are assumed to be ohmically contacted by the external metal leads with work-functions $5.6$ eV and $5.8$ eV for PdS$_2$ and NiS$_2$, respectively, and $5.3$ eV/$6.0$ eV for the $n$-type/$p$-type PtS$_2$. The monolayer 2D channel, with length $L_{\rm{ch}}$, is embedded in top and bottom SiO$_2$, with thickness $t_{\rm{ox}}=0.5$ nm.  

Sketches of the band-edge profiles of the PdS$_2$ and NiS$_2$ LH-FETs with semimetal source and drain are shown in Fig.~\ref{fig:Fig1}(d); the PtS$_2$ LH-FET with small gap source and drain is shown in Fig.~\ref{fig:Fig1}(e). The complete MLWF Hamiltonian describing the channel including source and drain regions feeds the open-boundary Schr\"{o}dinger equation, within the Non-equilibrium Green Functions (NEGF)~\cite{Datta00} formalism, that is self-consistently  solved with the electrostatics of the whole device (see Methods) ~\cite{VIDES,VIDESwww}.

In order to study the potential performance of the considered LH-FETs for logic applications, we set a supply voltage $V_{\rm{dd}} = 0.5$~V, and we simulate the transfer characteristics for a drain-to-source voltage  $V_{\rm{DS}}=V_{\rm{dd}}$. Those are shown in Fig. \ref{fig:Fig2} in semilogarithmic scale for channel lengths ranging from $L_{\rm ch}\simeq 5$~nm up to $L_{\rm ch}\simeq 10$~nm and for the three different materials. As it can be seen, the bilayer metallic source/drain regions in PdS$_2$ and NiS$_2$
lead to an ambipolar behavior. For channel lengths longer than $10$~nm, $I_{\rm off}$ is not further reduced, being the ambipolarity determined by the bandgap of the monolayer channel and the capability of the bilayer source to inject both kinds of carriers. In this regard, some improvement might be achieved following the back-gate over/underlapping strategy at the  source/drain discussed in Ref. \cite{Wu2018}. Here we have tuned the gate workfunction in the  PdS$_2$ and NiS$_2$ devices to $5.52$~eV  so to observe the crossover between $n$-type and $p$-type conduction at $V_{\rm{GS}}=0$~V.

The PtS$_2$ LH-FET is not ambipolar since it has semiconductor source and drain. In this case we show in Fig. \ref{fig:Fig2}(c) and (d) both the nFET and the pFET characteristics, where the gate workfunction is tuned to $5.28$~eV and $6.2$~eV, respectively, to obtain the current in the OFF state $I_{\rm off} = 100$ ~nA/$\mu$m at $V_{\rm GS}=0$~V, as required by the IRDS \cite{IRDS} for high performance applications where the OFF state corresponds to $V_{\rm GS}=0$~V and $V_{\rm{DS}}=\pm V_{\rm{dd}}$. The source and drain regions have a donor doping with molar fraction $N_{\rm D}=4.1\times 10^{-2}$ in the case of nFET, and acceptor molar fraction $N_{\rm A}=9.5\times 10^{-2}$ in the case of pFET. The channel is undoped for all devices.

\begin{figure} [t]
	\centering
	\includegraphics[width=0.6\columnwidth]{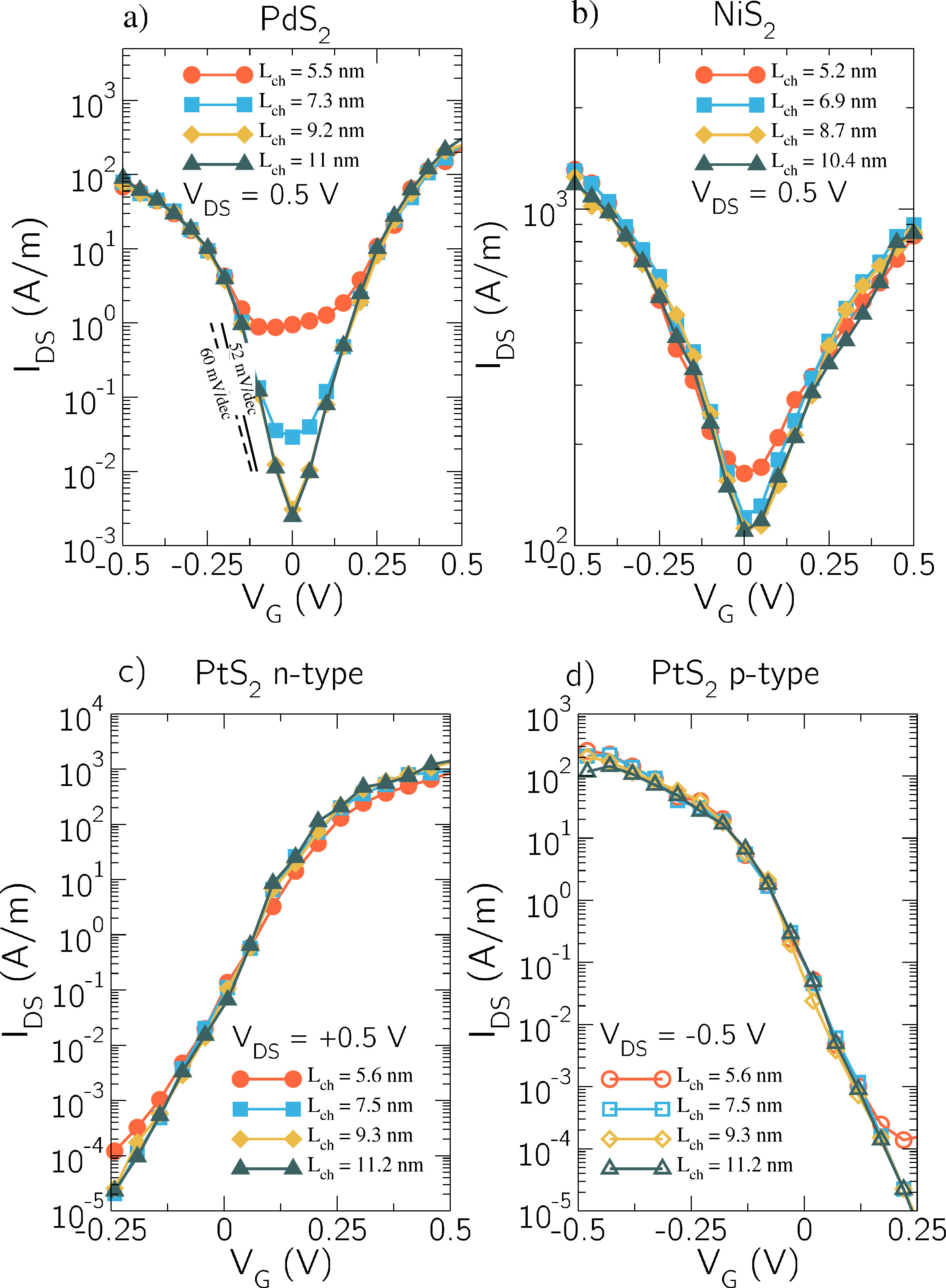}
	\caption{Transfer characteristic in semilogarithmic scale for the LH-FETs with PdS$_2$ (a), NiS$_2$ (b) and \emph{p}-type/\emph{n}-type PtS$_2$ (c)/(d) considering a channel length ranging from $\simeq5$~nm up to $\simeq10$~nm and drain-to-source voltage ${|V_{\rm DS}|=0.5}$~V. In the case of PdS$_2$ and NiS$_2$ an ambipolar behavior is observed due to the semimetallic source and drain. In the case of PtS$_2$ we consider both the nFET and the pFET. The channel is always undoped. A metal gate workfunction of $5.52$~eV is assumed in the  PdS$_2$ and NiS$_2$ devices to so to observe the crossover between $n$-type and $p$-type conduction at $V_{\rm{GS}}=0$~V. For the PtS$_2$ nFET and pFET, the gate workfunction is tuned to $5.28$~eV and $6.2$~eV, respectively, to set $I_{\rm off} = 100$ ~nA/$\mu$m at $V_{\rm{GS}}=0$~V.}
	\label{fig:Fig2}
\end{figure}

\begin{figure} [t]
	\centering
	\includegraphics[width=0.9\columnwidth]{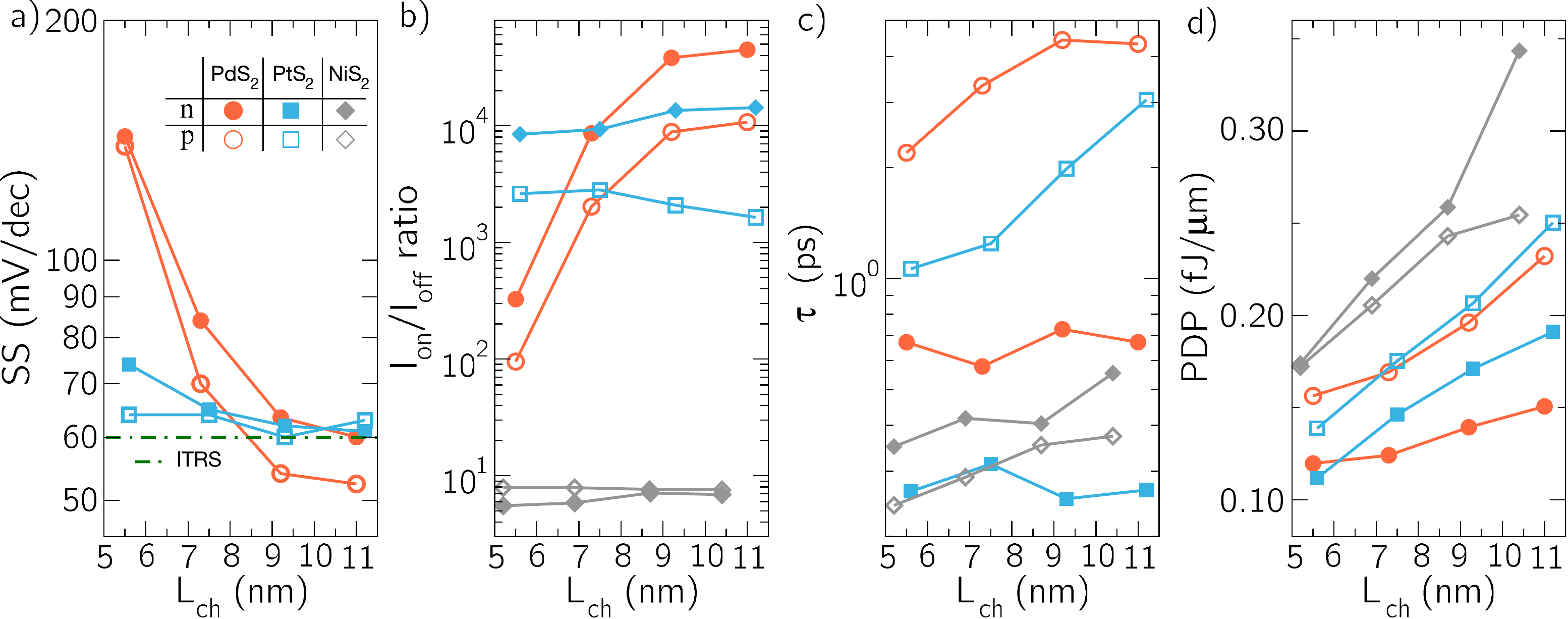}
	\caption{SS, $I_{\rm on}/I_{\rm off}$, $\tau$ and PDP as a function of the channel length, $L_{\rm ch}$, for the LH-FETs based on PdS$_2$, PtS$_2$, and NiS$_2$, evaluated according to the IRDS requirements for high-performance applications: $I_{\rm off}=100$nA/$\mu$m.}
	\label{fig:Fig3}
\end{figure}

Interestingly, the \emph{p}-type branch of the PdS$_2$ LH-FET $I_{\rm{DS}}-V_{\rm{GS}}$ curve exhibits a sub-maxwellian SS down to 52~mV/decade, as shown in Fig.~\ref{fig:Fig3}(a). The possibility to achieve sub-$60$ mV/decade SS in a thermionic FET has been subject of debate \cite{Cheung13, Solomon10}. In particular, in Ref. \cite{Solomon10} the authors argue that the $60$ mV/decade limit cannot be beaten in a single barrier device, although they eventually conclude that the role of the DOS can be essential to reverse this situation, as has already been experimentally demonstrated in the case of a graphene source \cite{Qiu18}.

\begin{figure} [t]
	{\includegraphics[width=0.8\columnwidth]{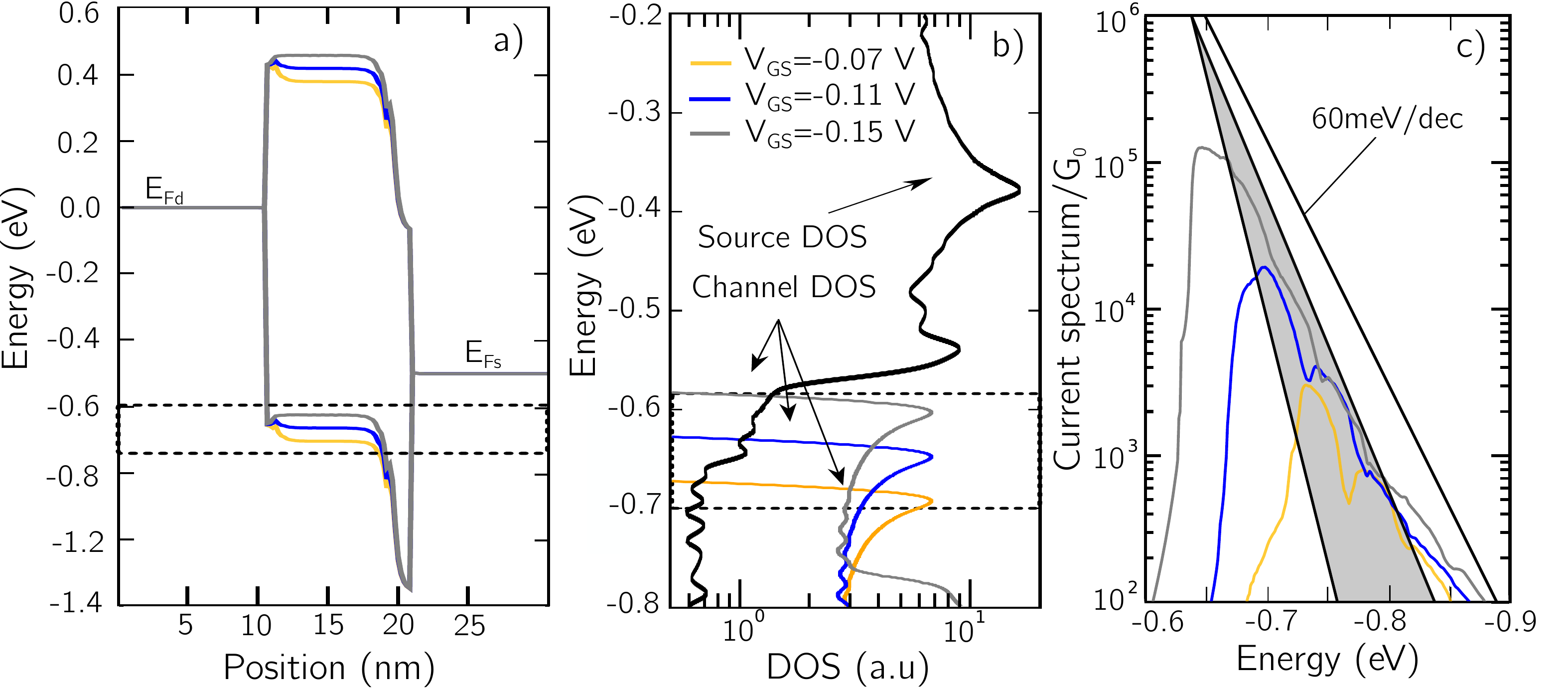}}
	\caption{a): Conduction and valence band edge profiles of the PdS$_2$ LH-FET for $V_{\rm{DS}}=-V_{\rm{dd}}$ and three different values of $V_{\rm{GS}} = -0.07, -0.11, -0.15$~V in the subthreshold region of the p-branch. b) Density of states of the source and of the channel for the corresponding values of $V_{\rm{GS}}$ obtained in a dense k-mesh grid, see Supporting Information, and after a Gaussian smoothing for the energy integration with $\sigma=10$~meV; c) Current spectrum as a function of energy. In shaded gray the range of slopes between $30$~mV/decade and $50$~mV/decade. The $60$~mV/decade Boltzmann limit is also plotted.}
	\label{fig:Fig4}
\end{figure}

We can discuss the specific mechanism in the PdS$_2$ LH-FET by considering the band edge profiles shown in 
Fig.~\ref{fig:Fig4}(a) for $V_{\rm{DS}} = V_{\rm{dd}}$ and for three different values of $V_{\rm{GS}}$ in the subthreshold region of the \emph{p}-type branch.
The Schottky barriers of drain-channel and source-channel junctions are calculated from first-principles (see Supplementary Information). 
A variation of $V_{\rm{GS}}$ modulates both the conduction and valence band edges in the channel and the transparency of the Schottky barrier between source and channel. The drop of the DOS in the source for energy below the source Fermi level (corresponding to $-0.5$~eV for the considered bias point) is clearly visible in Fig.~\ref{fig:Fig4}(b), (where the DOS is obtained from refined calculations applying, for visualization purposes, a Gaussian smoothing with $\sigma=10$~meV --see Supporting Information--) and is responsible for a sharper energy filtering than that provided by the Maxwell-Boltzmann tail of the occupation factor, and therefore for an SS lower than the so-called Boltzmann limit. This is even more apparent by considering the energy spectrum of the current ---\latin{i.e.} the current density per unit energy--- for different values of $V_{\rm{GS}}$ shown in Fig.~\ref{fig:Fig4}(c). The slope of the logarithm of the current spectrum as a function of energy would be exactly $-1/(k\rm{_B} T)$ for a constant DOS, corresponding to one decade every 60~meV at room temperature. Due to the non-constant DOS in the source, such slope is however not constant, and in one decade around an energy of $-0.7$~eV, where the DOS is dramatically quenched, it is in the range of 30~meV to 50~meV, marked as a shaded region in Fig.~\ref{fig:Fig4}(c). This more confined current spectrum is responsible for the sub-Maxwellian subthreshold swing of the \emph{p}-type branch transfer characteristics of the PdS$_2$ LH-FET.

Energy filtering is not observed in the $n$-branch because the DOS of the conduction band states in PdS$_2$ do not show a similar steep transition (see the Supporting Information for more details). The effect is neither observed in NiS$_2$ where the small gap of the monolayer semiconductor results in large interband tunneling currents and very poor SS (above 200~meV/decade and therefore not shown in Fig.~\ref{fig:Fig3}(a)). The PtS$_2$ LH-FET has a close to ideal Maxwellian SS of $60$~mV/decade for HP applications due to the dominant Fermi window tail that restricts the SS.

Other transistor figures of merit for digital electronics are considered in Fig.~\ref{fig:Fig3} following the IRDS specifications for high-performance (HP) applications. In particular, Fig.~\ref{fig:Fig3}(b) shows $I_{\rm on}/I_{\rm off}$ ratio, where $I_{\rm on}$ is the drain current in the ON state -- corresponding to $V_{\rm{GS}}=V_{\rm{DS}}=\pm V_{\rm{dd}}$ and $I_{\rm off}=100$~nA/$\mu$m is the current in the OFF state as defined by the IRDS for HP. To this purpose, the PdS$_2$ LH-FET exhibits the highest $I_{\rm on}/I_{\rm off}$ ratio for channel length close to 10~nm thanks to the lower SS just discussed, but not for shorter channel lengths (down to 5~nm), because of the high $I_{\rm off}$ due to ambipolarity and large source-to-drain tunneling. 
The situation  is worse for the NiS$_2$ LH-FET: its smaller monolayer bandgap ($0.47$~eV) leads to a very poor $I_{\rm on}/I_{\rm off}$ ratio.
The PtS$_2$ LH-FET, instead, has a semiconducting bandgap and negligible source-to-drain tunneling, and therefore exhibits high $I_{\rm on}/I_{\rm off}$ ratio for  $L_{\rm{ch}}$ down to 5~nm.

Relevant figures of merit for transistor performance in digital circuits are also the intrinsic delay time $\tau=(Q_{\rm on}-Q_{\rm off})/I_{\rm on}$ and the Power-Delay-Product PDP$=V_{\rm dd }\tau I_{\rm on}$, where $Q_{\rm on}$ and $Q_{\rm off}$ are the total mobile charge in the channel in the ON and OFF states, respectively (Fig. \ref{fig:Fig3}(c) and \ref{fig:Fig3}(d)). 
The nFETs based on PdS$_2$ and PtS$_2$ exhibit expected $\tau$ and PDP compliant with IRDS requirements for next technology nodes \cite{Iannaccone18,IRDS}, together with an $I_{\rm on}/I_{\rm off}$ ratio close to $10^4$ for channel lengths of at least 10~nm,  which implies acceptable stand-by power consumption for HP applications \cite{IRDS}. The pFETs have slightly worse PDP and $\tau$ than the nFETs, due to the smaller source DOS in the valence band -- apparent in Fig.~\ref{fig:Fig1}(a) -- which is responsible for a smaller $I_{\rm on}$, as can be seen in the asymmetric transfer characteristics of Fig.~\ref{fig:Fig2}. 

While the ambipolarity of PdS$_2$ and NiS$_2$ FETs spoils their use in low power (LP) applications, PtS$_2$ FETs  also satisfy the IRDS requirements to this purpose (\latin{i.e.} they reach an $I_{\rm off}=100$~pA/$\mu$m) achieving $I_{\rm on}/I_{\rm off}$ ratios above $10^6$ and $10^5$ for \emph{n}-type and \emph{p}-type FETs, respectively (see Supporting Information). For low stand-by power, SS is  spoiled in shorter channel lengths since the tunneling current becomes comparable to the pursued $I_{\rm off}$: for $L_{\rm ch}\approx 5-8$~nm it is in the range $80$-$100$~mV/decade for nFETs and $70$-$90$~mV/decade for pFETs, deviating considerably from the $60$ mV/decade observed in longer channels. Finally, the PDP is slightly lower and the intrinsic delay time slightly higher than the values obtained for HP (see Supporting Information for details).

\subsection{Resonant Tunneling Diode}
Finally, we have studied the operation of a 2D Resonant Tunneling Diode based on a PtS$_2$ LH (LH-RTD). 
We have considered bilayer PtS$_2$ drain and source regions with length $11.2$~nm. Two monolayer regions act as energy barriers confining a bilayer well. The barrier length is $L_{\rm b}=1.9$~nm and the length of the well $L_{\rm w}$ is varied from $1.9$~nm up to $5.6$~nm. The whole 2D channel is embedded in SiO$_2$.
Fig.~\ref{fig:Fig5}(a) shows the $I_{\rm DS}$ vs $V_{\rm{DS}}$ characteristics for $L_{\rm w}=1.9$~nm at room temperature. The current has a clear non-monotonic behavior exhibiting a pronounced negative differential resistance, due to resonant tunneling through quantized states in the 2D well. The local DOS in the channel is plotted in Fig.~\ref{fig:Fig5}(b) as a function of energy and of position along the device length ($y$) for $V_{\rm{DS}} = 0.25$~V and $0.5$~V, corresponding to the main peak and to the valley of the current-voltage characteristics.
Let us stress the fact that the bandstructure of different regions is fully considered in the calculation, but inelastic processes and heterojunction defects are not included. 

\begin{figure} [h!]
	\includegraphics[width=1.0\columnwidth]{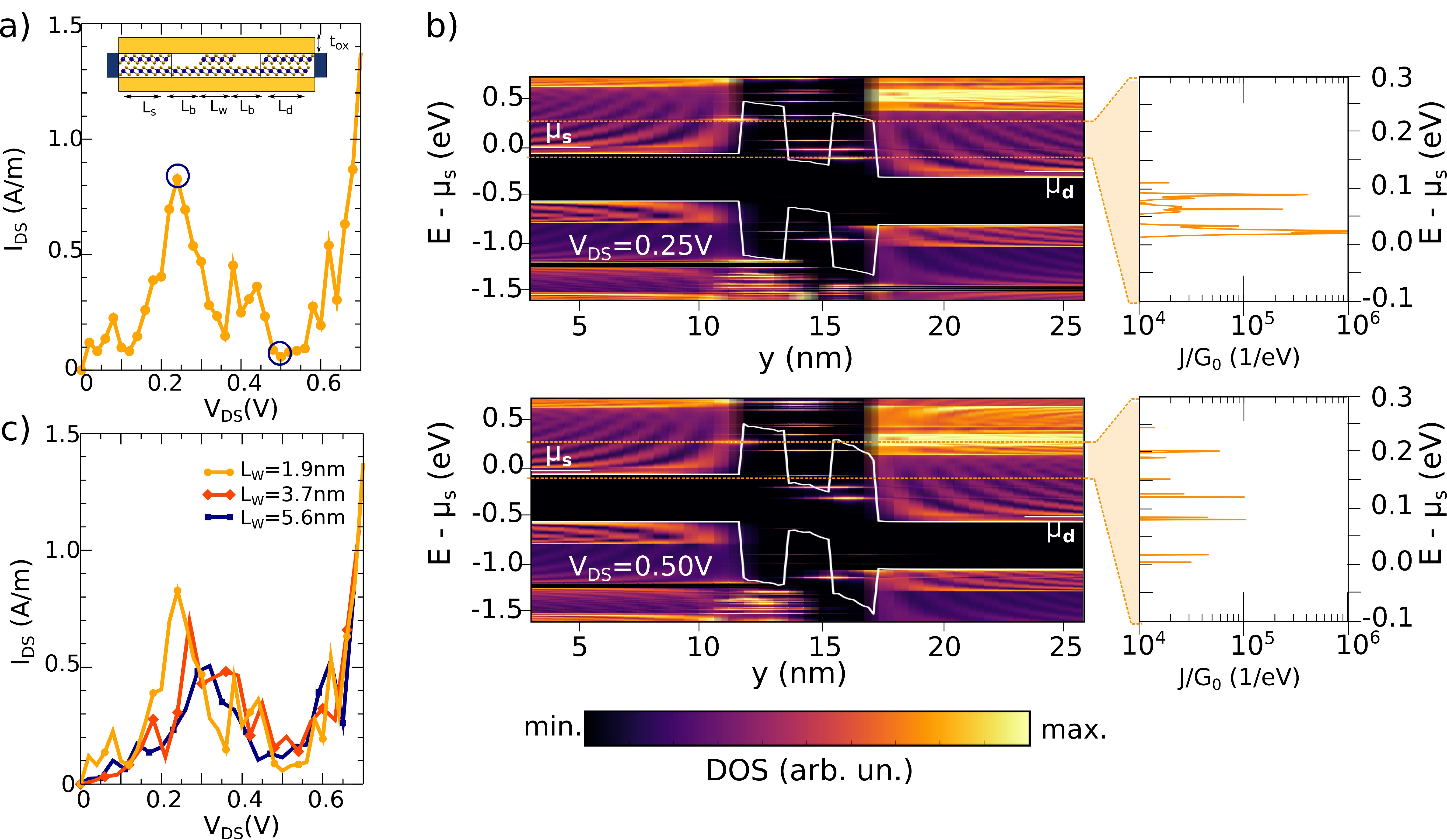}
	\caption{  a) $I_{\rm DS}$ vs $V_{\rm{DS}}$ characteristic of the LH-RTD based on PtS$_2$ with $L_{\rm w}=1.9$~nm. Inset: Schematic of the Lateral Heterostructure RTD b) Local DOS as a function of the energy  and the position along the device length, and current-density spectrum, normalized to the conductance quantum ($G_0=2q^2/\hbar$), for the RTD with a $L_{\rm w}=1.9$~nm, for $V_{\rm{DS}} = 0.25$~V and $0.5$~V. c) Current-voltage characteristics of the LH-RTD for different well lengths.}
	\label{fig:Fig5}
\end{figure}
The Fermi level at the source ($\mu_{\rm s}$) and drain ($\mu_{\rm d}$) leads are marked. We have super-imposed the conduction and valence band profiles to the local DOS colormaps. The alignment of $\mu_{\rm s}$ with these quantized energy levels results in resonances in the current spectrum density and is controlled by $V_{\rm{DS}}$. We have also explored different well lengths: $L_{\rm w}=1.9$~nm, $3.7$~nm and $5.6$~nm (Fig. \ref{fig:Fig5}(c)), observing that the NDR effect is preserved, although the position of the peak varies with the well length as so does the energy quantization, while the position of the valley is not modified as it depends on the height of the barrier limiting the thermionic emission. 

\section{Conclusion} 

We have shown that the strong dependence of the bandgap upon the number of layers of noble TMDs can be used to devise electron devices based on transport through lateral heterostructures of TMDs, such as LH-FETs and LH-RTDs.

We have used \latin{ab-initio} multiscale simulations to demonstrate that LH-FETs based on PdS$_2$ and PtS$_2$  can comply with IRDS performance requirements \cite{IRDS} for future integrated circuit technology for high performance digital applications. On the other hand, LH-FETs made of NiS$_2$ cannot meet such requirements, due to the low gap and ambipolar behavior. 

We have also predicted the steep (submaxwellian) subthreshold behavior of \emph{p}-type FETs based on PdS$_2$, due to the asymmetry of the bilayer PdS$_2$ DOS around the Fermi level, achieving SS=$52$~mV/decade. This is the first demonstration of this effect in an intrinsic 2D material, and can be further exploited for device design. It must be noted that in presence of electron-electron or electron-phonon scattering the behavior of the studied devices would be degraded with respect to the optimum ballistic condition assumed here, impacting directly on the achieved sub-Boltzmann slope. Moreover, the SS value for PdS$_2$ FETs is not expected to boost the device performance stunningly as compared to the conventional limit. However, and more importantly, the results presented here confirm that the $60$mV/decade limit can be beaten in 2D-based FETs, as it has been proved experimentally in graphene, encouraging the exploration of new 2D materials with sharper DOS and consequently steeper SSs.

Finally, we have also predicted the possibility of using 2D LHs to obtain a resonant tunneling diode, with a pronounced peak-to-valley ratio of the current-voltage characteristics, which is suitable for experimental observation.

\section{Methods}
Density Functional Theory as implemented in Quantum Espresso code has been employed to determine the electronic structure of PdS$_2$, NiS$_2$ and PtS$_2$. The crystal geometry of monolayer and bilayer 2D crystals is characterized by a 1T arrangement, with a layer of Pt/Pd/Ni atoms sandwiched between two atomic layers of S atoms. The atoms coordinates and lattice vectors has been obtained after \cite{Miro14b}, where a structural optimization of the unitary cell was performed.  We have considered $40$ \AA{} of vacuum in the direction orthogonal to the 2D layers to minimize spurious interactions between periodic repetitions of the cell. For the exchange-correlation functional, the local density approximation has been considered under the Perdew-Zunger \cite{PZ} parametrization within norm-conserving pseudopotentials. The energy cutoffs for charge density and wavefunction expansions have been set to $360$ Ry and $60$ Ry respectively. A Monkhorst$-$Pack $10\times5\times1$ $k$-mesh has been used for the Brillouin-zone integration and an energy convergence threshold of $10^{-6}$ eV in the iterative solution of the Kohn-Sham equations was ensured.  Additionally, an analysis of the Schottky barriers formed at the 2L/1L heterojunction has been performed (see Supplementary Information).

Maximally-Localized-Wannier-Functions (MLWFs) have been obtained by means of the Wannier90 code \cite{Wannier} for every material and stacking. For the change of basis, the same $k$-sampling of the Brillouin zone as in the DFT simulations has been used to compute the overlap matrices required to determine the MLWFs. $12$ bands around the fundamental gap have been considered and a threshold of $10^{-10}$ \AA{} has been set for the total spread change in the MLWFs in 20 consecutive iterations. The MLWFs band-structures have been calculated along the same path as in DFT, showing very good agreement (see Supplementary Information Fig. S1). The MLWF Hamiltonians of the 1L and 2L regions have been employed to build the total Hamiltonian of the lateral heterostructure following the procedure presented in Ref. \cite{Marian17}. 

The device simulations consist of the self-consistent solution of the open-boundary Schr\"{o}--dinger equation, within the Non-equilibrium Green Functions (NEGF)~\cite{Datta00} formalism, and the Poisson equation, for which we have used the open-source code NanoTCAD ViDES~\cite{VIDES,VIDESwww}. The construction of the Hamiltonian of the heterostructure from the Hamiltonians of the different regions/materials requires a careful treatment, with special attention to the mixing of the interface elements. We have followed the procedure we developed and validated in Ref. \cite{Marian17}. In particular, the off-diagonal elements connecting the 1L and 2L regions and determining their coupling have been assumed to be equal to those of the monolayer region. We have tested different alternatives for the off-diagonal coupling elements at the interface (see Supplementary Information) observing little variation in the device behavior. This mixing procedure provides the best results in terms of robustness in the convergence, preserving computational accuracy as compared to \latin{ab-initio} simulations. For all devices we have considered operation at temperature of 300~K.

\begin{acknowledgement}

Authors gratefully acknowledge the support from the European Commission through the Graphene Flagship Core 2 (contract n. 785219) and through the QUEFORMAL h2020 Project (contract n. 829035).
E.G. Marin also acknowledges Juan de la Cierva Incorporaci\'on IJCI-2017-32297 (MINECO/AEI)

\end{acknowledgement}


\bibliography{biblio}

\clearpage
\newpage

\section{\Huge Supplementary Information} 

\section{Hamiltonian of the lateral heterostructure} 

\vspace{-0.2cm}
In order to build the heterostructure Hamiltonian, one must take notice of the off-diagonal
elements connecting the monolayer and the bilayer regions. In particular, in the Hamiltonian along
the transport direction each block represents a
(\#wannier centers $\times$ \#wannier centers) matrix, and the number of blocks along a row is 
constrained by the Monkhorst-Pack grid. The blocks in the off-diagonal elements connect a row of the
bilayer region to a column of the monolayer region, i.e. they connect the last cell
of the source and the first of the channel (and equivalently at the drain end).

\begin{figure} [h!]
	\centering
	\includegraphics[width=0.28\columnwidth]{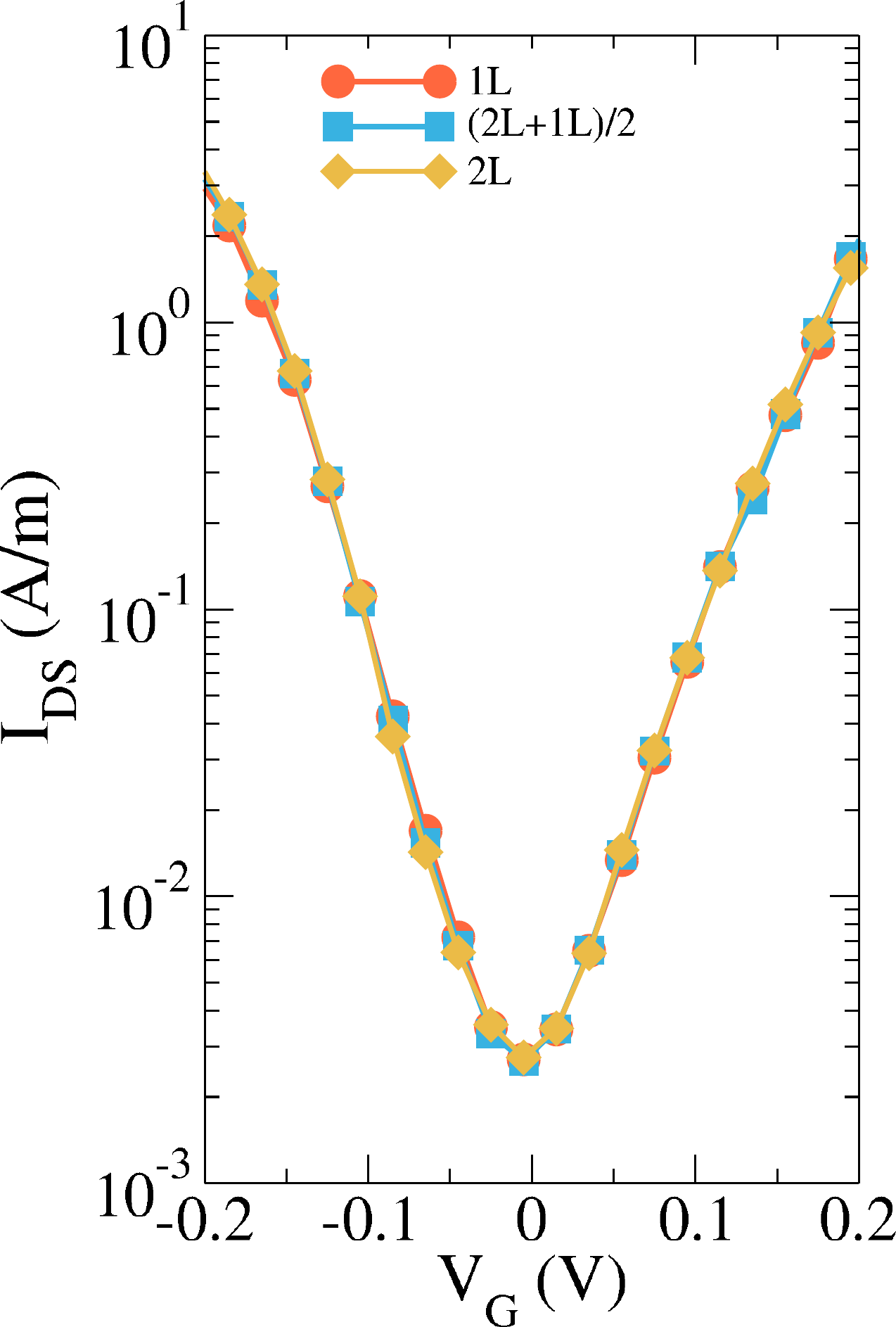}
	\caption{Transfer characteristic for the bilayer-monolayer-bilayer PdS$_2$ lateral-heterostructure FET with a channel length $9.2$~nm
		assuming different coupling between the bilayer and monolayer regions.}
	\label{fig:Fig2}
\end{figure} 

\noindent We have tested how different couplings between the monolayer and bilayer affect to the device. Figure \ref{fig:Fig2} shows the transfer response of the PdS$_2$ LH-FET with $9.2$-nm long channel, comparing three mixing strategies: 1) the one employed in the manuscript, where the off diagonal elements are assumed to be equal to those of the monolayer region (red circles) 2) a mean of the 1L and 2L coupling values (blue squares), and 3) the bilayer coupling values (yellow diamonds). We observe that regardless the values assigned to the off-diagonal coupling parameters the transfer characteristic change very little and the SS$<$60mV/decade is conserved.

\clearpage
\newpage

\section{Band structure}

\begin{figure} [h!]
	\centering
	\includegraphics[width=0.5\columnwidth]{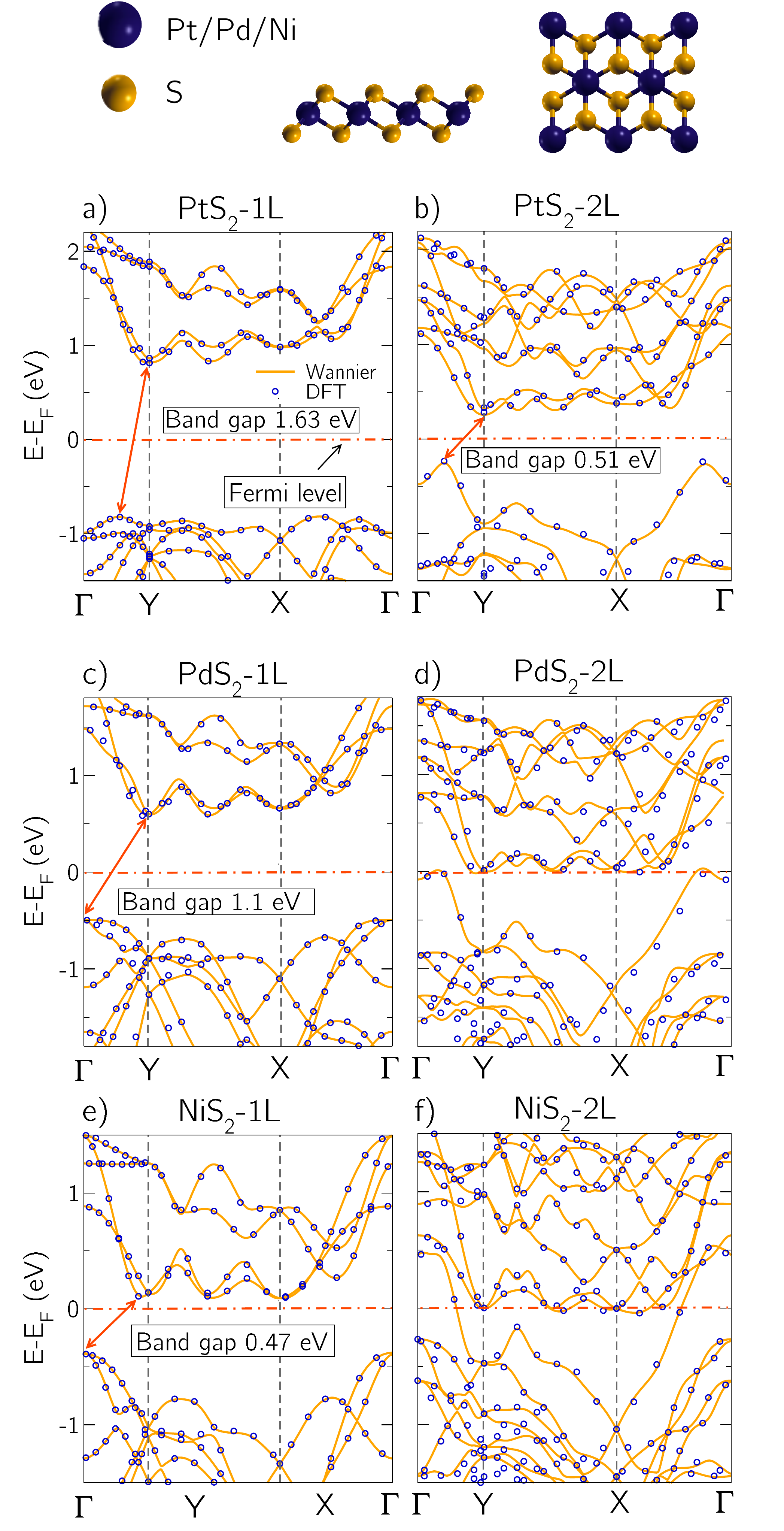}
	\caption{Top and lateral views of the 1T crystal structure of monolayer and bilayer PtS$_2$, PdS$_2$, and NiS$_2$. They are characterized by a 1T structure,  with a layer of Pt/Pd/Ni atoms sandwiched between two atomic layers of S. Electronic band-structure on a highly symmetric path along the Brillouin zone as computed with Density Functional Theory calculations (symbols) and with Maximally Localized Wannier Functions (lines).}
	\label{fig:FigS1}
\end{figure}
\clearpage
\newpage

\section{Schottky Barrier}

To deepen in the analysis of the 2L-1L interface and obtain accurate information on the Schottky Barrier (SB) formation and the alignment of the bandstructures of the two regions, we have performed an analysis of a complete 2L-1L heterojunction. First principles calculations using the Quantum Espresso suite \cite{QE} (see Methods) have been performed for the 2L-1L  heterojunction. The profile of the vacuum level and the potential profile with respect to the Fermi level, $E_{\rm F}$, has been calculated taking into account the formation of dipoles but neglecting the presence of defects. Following the same methodology as in \cite{Katagiri16}, we have extracted values for the SBs of $0.43$ eV, $0.26$ eV and $0.41$ eV for the PdS$_2$, NiS$_2$ and PtS$_2$ lateral heterostructures, respectively. These values have later been considered in the construction of the Hamiltonian for the device simulations.

\clearpage
\newpage

\section{Current spectrum for the \emph{p}-type PdS$_2$ LH-FET}

\begin{figure} [h!]
	\centering
	\includegraphics[width=0.5\columnwidth]{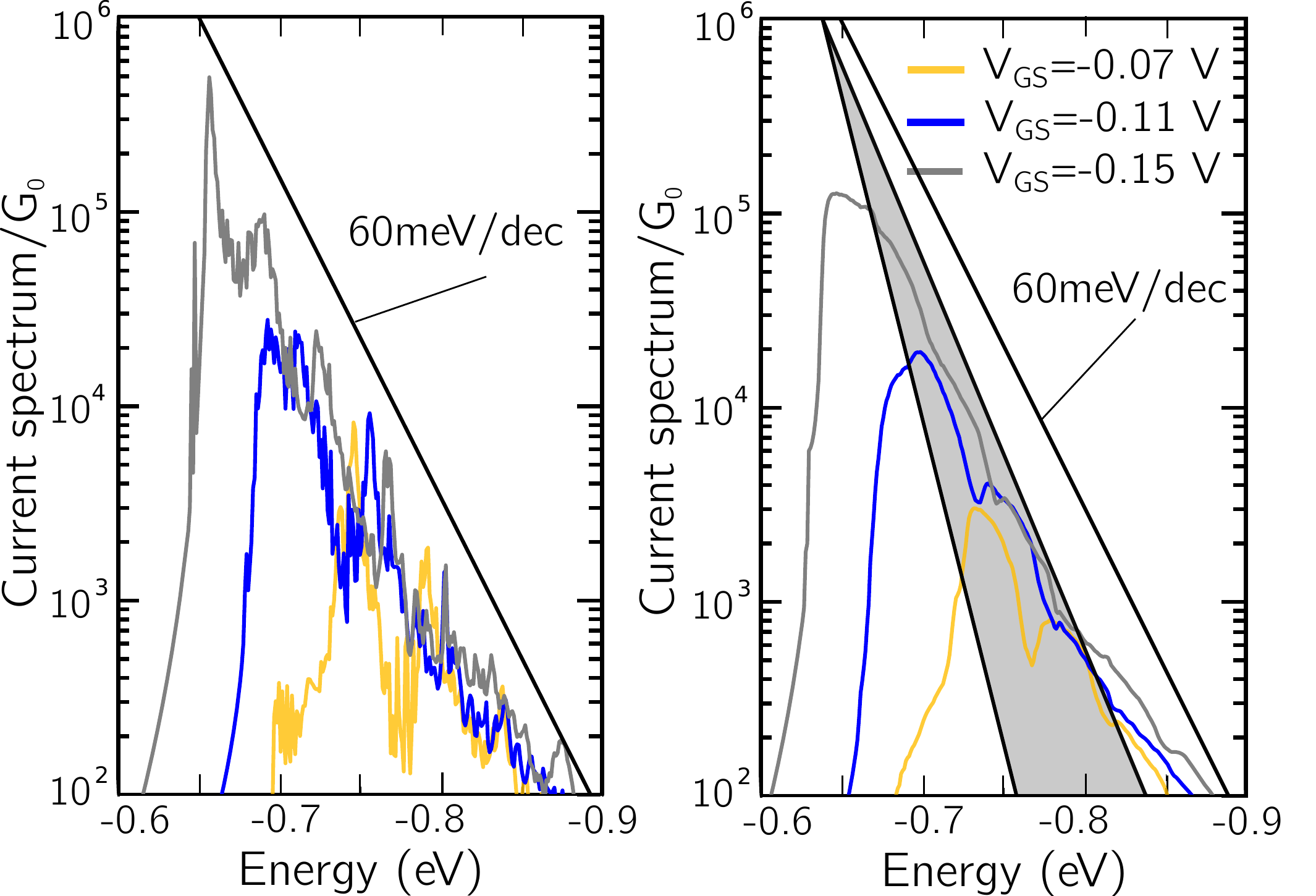}
	\caption{(Left) Current spectrum for the $L_{\rm ch}=9.2$~nm  PdS$_2$ LH-FET for the bias points in the p-branch where the sub-maxwellian SS is achieved. (Right) In order to appreaciate better the slope of the current spectrum vs. energy we low-pass filtered the current using an average energy window of $5$ meV. In shaded gray the range of slopes between $30$ mV/decade and $50$ mV/decade. The $60$ mV/decade Boltzmann limit is also plotted.}
	\label{fig:FigS3}
\end{figure}
\clearpage
\newpage

\section{Current spectrum for the \emph{n}-type PdS$_2$ LH-FET}

\begin{figure} [h!]
	\centering
	\includegraphics[width=0.3\columnwidth]{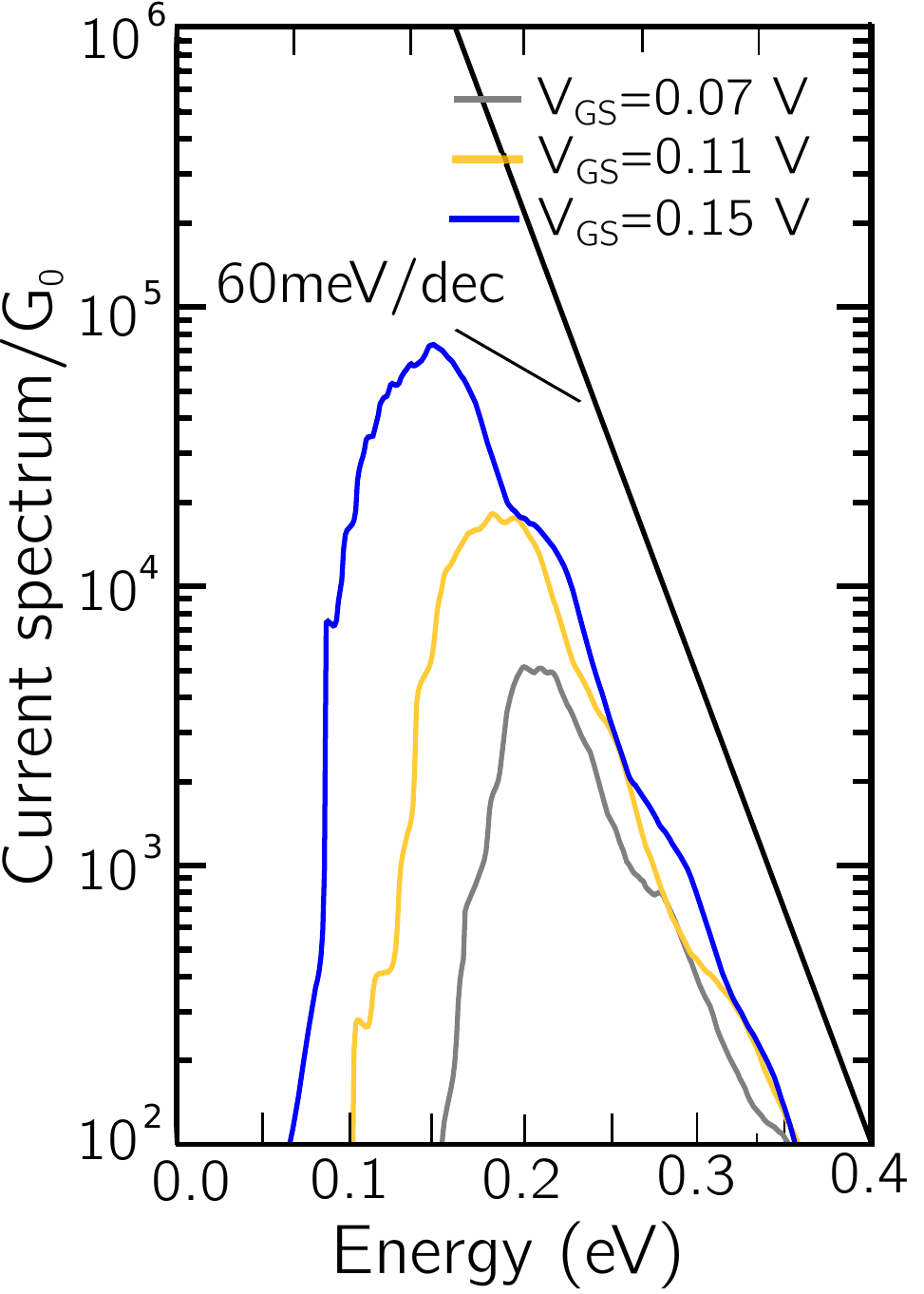}
	\caption{Current spectrum as a function of energy for the $L_{\rm ch}=9.2$~nm  PdS$_{2}$ LH-FET for three different values of $V_{\rm GS} = 0.07, 0.11, 0.15$ V in the subthreshold region of the \emph{n}-type branch. The $60$ mV/decade Boltzmann limit is also plotted. In order to
		appreaciate better the slope of the current spectrum vs. energy we low-pass filtered the spectrum using an average energy window of $5$ meV.}
	\label{fig:Fig5}
\end{figure} 

\clearpage
\newpage

\section{Figures of merit of the PtS$_2$ LH-FET for low stand-by power applications}

\begin{table} [h!]
	\begin{center}			
		\begin{tabular}{|c||c|c|c|c||c|c|c|c|}
			\hline
			\multirow{3}{*}{}  & \multicolumn{4}{c||}{\emph{n}-type}  &  \multicolumn{4}{c|}{\emph{p}-type} \\
			\hline
			\L$_{\text{ch}}$ & SS & I$_{\text{on}}$/I$_{\text{off}}$  & $\tau$ & PDP  & SS & I$_{\text{on}}$/I$_{\text{off}}$  & $\tau$ & PDP \\
			\ (nm) & (mV/dec) &  & (ps) & (fJ/$\mu$m)   & (mV/dec) &  & (ps) & (fJ/$\mu$m)  \\
			\hline
			5.6     & 108 & $ 1.1 \cdot 10^{6}$  & 1.15  & 0.04    & 90 & $3.4 \cdot 10^{5}$ & 6.1 & 0.10 \\
			
			7.5     & 80  & $3.6 \cdot 10^6$  & 0.49  & 0.09   & 73 & $7.5 \cdot 10^{5}$  & 3.4  & 0.13 \\
			
			9.3    & 70  & $3.9 \cdot 10^6$  & 0.49  & 0.10   & 60 & $7.5 \cdot 10^5$ &  3.8 & 0.14 \\
			
			11.2    & 63  & $4.5 \cdot 10^6$  & 0.47  & 0.11   & 60 & $6.3 \cdot 10^5$ & 4.8 & 0.15 \\				
			\hline
		\end{tabular}
		\caption{\label{Tab1} 
			{Figures of merit for several channel lengths of the PtS$_2$ for low-power applications. Differently from Figure 3 in the main text (evaluated for high-perfomance applications) the IRDS low-power specifications sets $I_{\rm off}=100$~pA$/\mu$m.}}
		\label{tab1}
	\end{center}
\end{table}	

\clearpage
\newpage

\section{Density of states calculation for the 1L and 2L PdS$_2$}

\begin{figure} [h!]
	\centering
	\includegraphics[width=0.3\columnwidth]{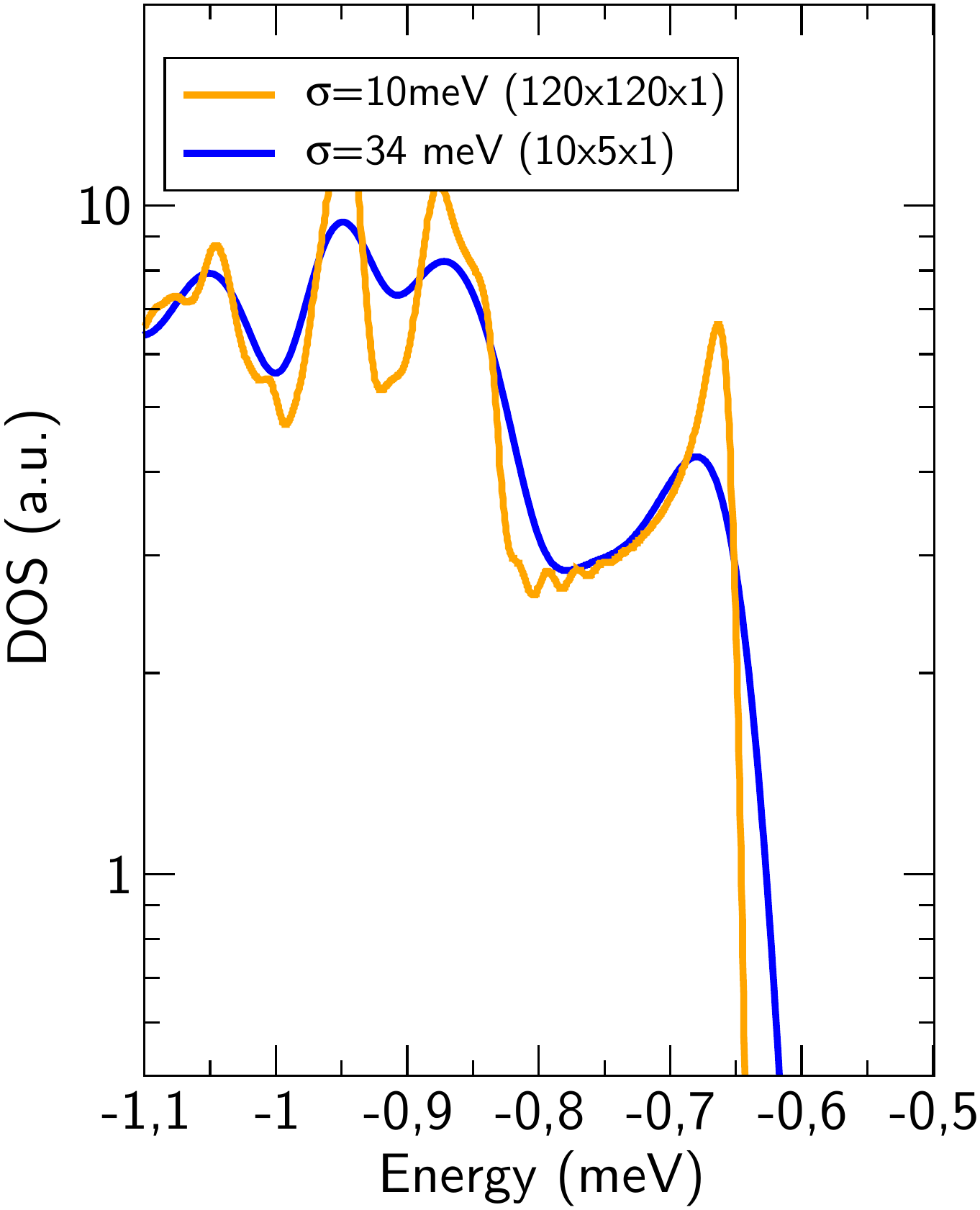}
	\includegraphics[width=0.3\columnwidth]{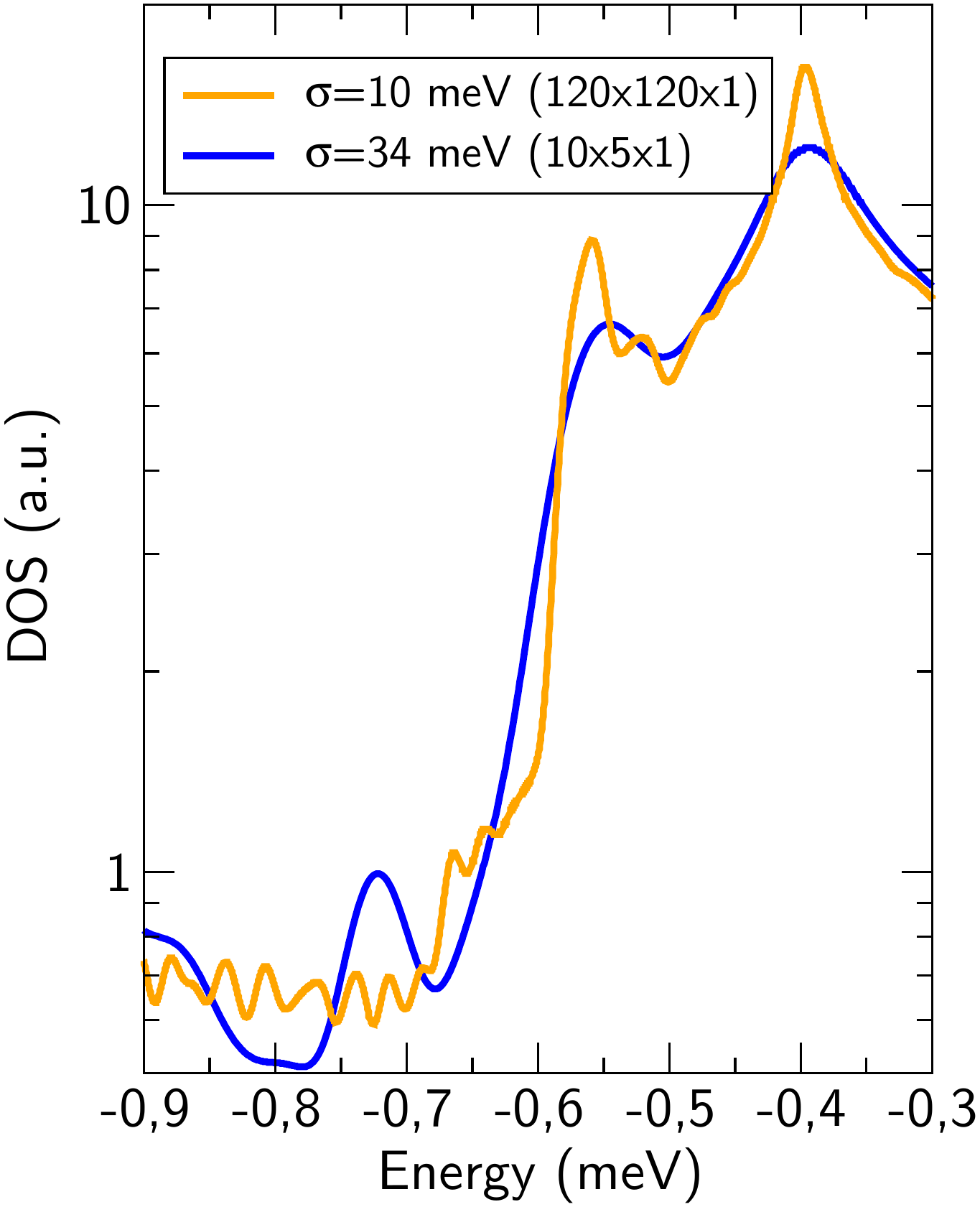}
	\caption{DOS of the monolayer (left) and bilayer (right) PdS$_2$  in a $10\times5\times 1$ Monkhorst-Pack grid with $0.5$~meV of energy resolution and considering a Gaussian smoothing for the energy integration $\sigma=34$meV (blue) and in a $120\times120\times 1$ Monkhorst-Pack grid with energy resolution $0.5$~meV and after a Gaussian smoothing with $\sigma=10$~meV (orange).}
	\label{fig:DOS_SI}
\end{figure} 

\clearpage
\newpage

\end{document}